\documentclass[12pt]{article}

\usepackage{graphics,epsfig}
\usepackage{amsbsy}
\usepackage{amsfonts}
\usepackage{amsmath}
\usepackage{graphicx}

\def\Xint#1{\mathchoice
{\XXint\displaystyle\textstyle{#1}}%
{\XXint\textstyle\scriptstyle{#1}}%
{\XXint\scriptstyle\scriptscriptstyle{#1}}%
{\XXint\scriptscriptstyle\scriptscriptstyle{#1}}%
\!\int}
\def\XXint#1#2#3{{\setbox0=\hbox{$#1{#2#3}{\int}$ }
\vcenter{\hbox{$#2#3$ }}\kern-.6\wd0}}

\def\dashint{\Xint-}

\def\res#1#2{\mbox{Res}\left[#1,#2\right]}
\def\ppint{\dashint}

\def\id{\protect{{1 \kern-.28em {\rm l}}}}

\newcommand{\beq}{\begin{equation}}
\newcommand{\eeq}{\end{equation}}
\newcommand{\beqr}{\begin{displaymath}}
\newcommand{\eeqr}{\end{displaymath}}
\newcommand{\beqa}{\begin{eqnarray}}
\newcommand{\eeqa}{\end{eqnarray}}
\newcommand{\beqar}{\begin{eqnarray*}}
\newcommand{\eeqar}{\end{eqnarray*}}

\def\k{\kappa}

\def\p{{\partial}}
\def\nn{\nonumber}

\newcommand{\ads}[1]{\mbox{${AdS}_{#1}$}}
\newcommand{\adss}[2]{\mbox{$AdS_{#1}\times {S}^{#2}$}}

\def\dalemb#1#2{{\vbox{\hrule height .#2pt
        \hbox{\vrule width.#2pt height#1pt \kern#1pt
                \vrule width.#2pt}
        \hrule height.#2pt}}}

\def\half{\frac{1}{2}}
\let\a=\alpha \let\b=\beta \let\g=\gamma \let\d=\delta \let\e=\epsilon
\let\z=\zeta  \let\th=\theta  \let\k=\kappa
\let\l=\lambda \let\m=\mu \let\n=\nu \let\x=\xi \let\p=\pi 
\let\s=\sigma \let\t=\tau   \let\c=\chi 
 \let\vep=\varepsilon
\let\w=\omega      \let\G=\Gamma \let\D=\Delta \let\Th=\Theta \let\L=\Lambda
 \let\P=\Pi \let\S=\Sigma  
\let\C=\Chi \let\W=\Omega
\let\la=\label \let\ci=\cite 
  
\def\nn{\nonumber} \def\bd{\begin{document}} \def\ed{\end{document}}
\def\ds{\documentstyle} \let\fr=\frac \let\bl=\bigl \let\br=\bigr
\let\Br=\Bigr \let\Bl=\Bigl
\let\bm=\bibitem
\let\na=\nabla
\def\tU{{\widetilde U}}
\let\pa=\partial \let\ov=\overline
\def\ie{{\it i.e.\ }}
\newcommand{\be}{\begin{equation}}
\newcommand{\ee}{\end{equation}}
\def\ba{\begin{array}}
\def\ea{\end{array}}
\def\ft#1#2{{\textstyle{{\scriptstyle #1}\over {\scriptstyle #2}}}}
\def\fft#1#2{{#1 \over #2}}
\def\F#1#2{{ F_{#1}^{(#2)} }}
\def\cF#1#2{{ {\cal F}_{#1}^{(#2)} }}

\def\={\, =\, }
\def\+{\, +\, }
\def\-{\, -\, }

\def\R{{\bf R}}
\def\sst#1{{\scriptscriptstyle #1}}
\def\oneone{\rlap 1\mkern4mu{\rm l}}
\def\e7{E_{7(+7)}}
\def\td{\tilde}
\def\wtd{\widetilde}
\def\im{{\rm i}}
\newcommand{\ho}[1]{$\, ^{#1}$}
\newcommand{\hoch}[1]{$\, ^{#1}$}
\newcommand{\bea}{\begin{eqnarray}}
\newcommand{\eea}{\end{eqnarray}}
\newcommand{\ra}{\rightarrow}
\newcommand{\lra}{\longrightarrow}
\newcommand{\Lra}{\Leftrightarrow}
\newcommand{\ap}{\alpha^\prime}
\newcommand{\bp}{\tilde \beta^\prime}
\newcommand{\cB}{{\cal B}}
\newcommand{\cO}{{\cal O}}
\newcommand{\vecx}{\vec{x}}
\newcommand{\vecy}{\vec{y}}
\newcommand{\vecp}{\vec{p}}
\newcommand{\vecq}{\vec{q}}
\newcommand{\tr}{{\rm tr} }
\newcommand{\Tr}{{\rm Tr} }

\newcommand{\cL}{{\cal L}}
\newcommand{\cA}{{\cal A}}
\newcommand{\cD}{{\cal D}}
\def\sst#1{{\scriptscriptstyle #1}}

\def\ve{\varepsilon}
\def\vf{\varphi}
\def\F{\Phi}
\def\wg{\wedge}

\def \foot {\footnote}
\def \bi{\bibitem}

\def \tr {{\rm tr}}
\def \ha {{1 \over 2}}
\def \td {\tilde}
\def \ci{\cite}
\def \N {{\mathcal N}}
\def \ww {\Omega}
\def \const {{\rm const}}
\def \ss {\sum_{i=1}^3 }
\def \t {\tau}
\def\S{{\mathcal S} }
\def \nn {\nu}
\def \XX {{\rm X}}

\def \lra {\leftrightarrow}
\def \vom {{\bar \omega}}
\def \E {{\mathcal  E}} \def \J {{\mathcal  J}}
\def \YY {{\rm Y}}

\def \d {\del}
\def \rJ {{J}}
\def \sms {sigma models\ }
\def \sm {sigma model\ }
\def \L {\Lambda}
\def \gl {\ell}
\def \tr {{\rm tr\ }}
\def\z{\zeta}
\def\zi{\zeta_1}
\def\zii{\zeta_2}
\def\K{\mbox{K}}
\def\eE{\mbox{E}}   \def \vt {\vartheta}
\def \vr {\varrho}
\def \wup {w}

\def\dg{\dagger}
\def\a{\alpha}
\def\b{\beta}
\def\e{\varepsilon}
\def\p{\phi}
\def\ap{\alpha^\prime}
\def\I{{\cal I}}

\def\R{{\bf R}}
\def\Z{{\bf Z}}
\def\C{{\bf C}}
\def\P{{\bf P}}
\def\xb{{\bar X}}
\def\Tr{{\rm  Tr}}
\def\tr{{\rm  tr}}

\def \del{\partial}
\def \a {\alpha}
\def \aa {{\a'}}
\def\g{\gamma}
\def\s{\sigma}
\def\z{\zeta}
\def\zi{\zeta_1}
\def\zii{\zeta_2}
\def\ov{\over}

\def\I{{\cal I}}
\def\J{{\mathcal J}}
\def \ok {{1\ov \k}}
\def\LL{{\mathcal L }}
\def \jL {{J}}
\def \om {\omega}
\def \cL {{\mathcal L}} \def \cH {{\mathcal H}}
\def\E{{\mathcal E}}
\def\w{\omega}
\def\b{\beta}
\def\l{\lambda}
\def\eps{\epsilon}
\def\vep{\varepsilon}
\def \De {{\mathcal D}}

 \def \cV {{\cal V}}
\def  \Jt {  {J}_{\rm tot}    }

\def \k {\kappa}
\def\foot{\footnote}
\def \four{{\textstyle {1\ov 4}}}
 \def \third { \textstyle {1\ov 3
}}
\def\det{\hbox{det}}
\def \ci {\cite}

\def \foot {\footnote}
\def \bi{\bibitem}

\def \tr {{\rm tr}}
\def \ha {{1 \over 2}}
\def \tid {\tilde}
\def \vv {{\rm v}}
\def \tl {{\tilde \l}}
\def \XX {{\rm X}}
\def \ta {{\tilde \a}}
\def \fo { {1\ov 4}}
\def \ep {\epsilon}
\def \inti {{\int^{2\pi}_0 {d \sigma \ov 2 \pi}}}

\def \d {\partial}
\def \K {{\rm S}}
\def \el {\ell}
\def \Tr {{\rm Tr}}
\def \P {\Phi}
\def \l  {\lambda}
\def \tl {{\tilde \l}}
\def \bl {{\tilde \l}}
\def \const {{\rm const}}
\def \V {v}

\def \bv {v^*}
\def \vv {{\rm v}}
\def \LL {{\mathcal L}}
\newcommand{\PV}[1]{P_{\!\!_{V_{#1}}}}

\def \bL {\ell}
\def \M {{\mathcal M}}
\def \N {{\mathcal N}}
\def \S {{\rm S}}
\def \vn {\vec n}
\def \tl {\td \l}
\def \td {\tilde}
\def \Prod {\Pi}
\def \O {{\mathcal O}}
\def \Q {{\rm  Q}}
\def \D {\Delta}
\def \N {{\mathcal N}}
\def\tN{{\tilde N}}

\def \m {\mu}
\def \vs {\vec \s}
\def \ie {i.e.}

\def \cD {{\cal D}}

\def  \le  {\l_{\rm eff}}

\def \rS {{\rm S}}
\def\as{{\a}}
\newcommand{\bra}[1]{\mbox{$\langle #1 |$}}
\newcommand{\ket}[1]{\mbox{$| #1 \rangle$}}

\newcommand{\Gg}{G}

\newcommand{\auth}{AUTHORS}

\def\thb{\bar{\theta}}
\def\Thb{\bar{\Theta}}
\def\barp{\bar{p}}
\def\barq{\bar{q}}
\def\barc{\bar{c}}
\def\bard{\bar{d}}
\def\e{\epsilon}

\def \bi{\bibitem}
\def \la {\label}

\def \l {\lambda}
\def\foot{\footnote}
\def \tl  {{\tilde \l}}
\def \sql {{\sqrt \l}}
\def \adss {$AdS_5 \times S^5$\ }
\newcommand{\rf}[1]{(\ref{#1})}
\def \ov {\over}

\def\th{\theta}
\def\Th{\Theta}
\def\vth{\vartheta}

\def\vth{\vartheta}

\def\ra{\rightarrow}
\def\N{{\cal N}}
\def\F{{\cal F}}
\def\cc{\circ}
\def\eqv{\equiv}

\def\ni{\noindent}
\def \ha{{1\ov 2}}
\def \bw {{\rm w}}

\def\r{{\rm r}}

\def \cT {{\cal T}}
\def \no {\nonumber}

\def \J {\mathcal{J}}
\def \del {\partial}

\def \bps {{\bar \psi}}
\def \sqbl {\sqrt{\bar \lambda}}

\def\dF{\dot{F}}
\def\dG{\dot{G}}
\def\df{\dot{f}}
\def \E {{\cal E}}

\def \S {{\cal S}}
\def \J {{\cal J}}

\def\ms{\mathcal{S}}
\def\mj{\mathcal{J}}
\def\soj{\fr{\ms}{\mj}}
\def \R {{\bf R}}
\def \om {\omega}
\def \tH {\widetilde H}
\def \bE {\bar E}
\def \x {{\cal X}}

\def \hV {{\hat V}}

 \def \bb {\bar \beta}
\def \W {{\cal E}}

\def \bi{\bibitem}
\def \la {\label}

\def \l {\lambda}
\def\foot{\footnote}
\def \tl  {{\tilde \l}}
\def \sql {{\sqrt \l}}
\def \sqtl {{\sqrt {\tilde \l}}}

\def \HH {{\rm E}}
\def \cS {{\cal S}}
\def \cL {{\cal L}}

\def \adss {$AdS_5 \times S^5$\ }

\def \D {\Delta}
\def \thet {\theta}
 \def \t {\tau}
 \def \p {\phi}
 \def \r {\rho}
 \def \rN {{\rm N}}
 \def\tw{{\tilde w}}
 \def\hJ{{J}}
 \def\hw{{w}}
 \def\hl{{\lambda}}
 \def\hth{{\theta}}
 \def\NN{{\cal N}}
 \def \bv {{ \bar w}}
\def \vn {{\vec n}}
\newcommand{\sfrac}[2]{{\textstyle\frac{#1}{#2}}}
\def \bl {{ \bar \lambda}}
\def \bp {{\bar p}}
\def \bu {{\bar u}}
\def \sha {\sfrac{1}{2}}
\def \w {\omega}
\def \ov {\over}
\def \vl { \vec \ell}
\def \varpi {{\rm w}}
\def \OO {{\cal O}}
\def \bG {\bar \G}

\def \c {\gamma}

\def \ss {{\rm s}}

\def \ve {\varepsilon}
\def \pa{\partial}
\def \I {{\cal I}}
\def \LL {{\cal L}}
\def \ep {\epsilon}
\def \R {{\rm R}}
\def \tilt {{\tilde t}}

\def\pic #1#2{\hbox{\lower#1pt\hbox{~\mbox{\epsfxsize=20truemm \epsffile{#2}}}}}

\def\pic #1#2#3{\hbox{\lower#1pt\hbox{~\mbox{\includegraphics[scale=#3]{#2}}}}}

\def \bt {\bar\theta}
\def \te {\theta}

\def \cc {{\rm f}}
\def \d {\delta}

\def \cL {{\cal L}}
\def \S  {{\cal S}}
\def \pp {{q}}
\def \vt {\vartheta}
\def \mm {{\cal  \ell}}
\def \Z {{\cal Z}}
\def \pa {\partial}
\def \C {{\cal C}}
\def \be {\bea}
\def \ee {\eea}
\def \c {\gamma}  \def \d {\delta}
\def \eps {\epsilon}

\def \bp {\begin{pmatrix}}  \def \ep {\end{pmatrix}}

 \def \T {{\cal T}}

\def \bp {\begin{pmatrix}}  \def \epm {\end{pmatrix}}
\def \ha {{\textstyle{1 \ov 2}}}

\begin{document}
\overfullrule=0pt
\parskip=2pt
\parindent=12pt
\headheight=0in \headsep=0in \topmargin=0in \oddsidemargin=0in

\vspace{ -3cm} \thispagestyle{empty} \vspace{-1cm}

\vspace{ -3cm} \thispagestyle{empty} \vspace{-1cm}

\begin{center}
 \vspace{2cm}
{\Large\bf
Spiky strings in Bethe Ansatz at strong coupling}

 \vspace{.5cm} {
  M. Kruczenski$^{}$\footnote{markru@purdue.edu}
 and A.
 Tirziu$^{}$\footnote{atirziu@purdue.edu
 }}\\
 \vskip 0.3cm

{\em
$^{}$Department of Physics, Purdue  University,\\
W. Lafayette, IN 47907-2036, USA.\\
}

\end{center}

 \begin{abstract}

We study spiky string solutions in $AdS_3 \times S^1$ that are characterized by two spins $S,J$
as well as winding $m$ in $S^1$ and spike number $n$. We construct explicitly two-cut solutions by using the $SL(2)$ asymptotic Bethe Ansatz equations  at leading order in strong coupling. Unlike the folded spinning string, these solutions have asymmetric distributions of Bethe roots. The solutions match the known spiky string classical results obtained directly from string theory for
arbitrary semiclassical parameters, including $\mathcal{J}=0$ and any value of $\mathcal{S}$, namely short and long strings.
At large spins and winding number the string touches the boundary, and we find a new scaling limit with the energy given as
$
\mathcal{E}-\mathcal{S}=\frac{n}{2 \pi} \sqrt{1+ \frac{4 \pi^2}{n^2}\bigg(\frac{\mathcal{J}^2}{\ln^2 \mathcal{S}}+\frac{m^2}{\ln^2 \mathcal{S}}\bigg)}\ln \mathcal{S}
$. This is a generalization of the known scaling for the folded spinning string.

\end{abstract}
\newpage

\renewcommand{\theequation}{1.\arabic{equation}}
 \setcounter{equation}{0}

\setcounter{equation}{0} \setcounter{footnote}{0}
\setcounter{section}{0}

\def \ads {$AdS_5$ \ }

\section{Introduction}

Asymptotic $SL(2)$ Bethe Ansatz equations (ABA) played a crucial role recently in matching the energy of the free string with the
dimension of the corresponding operators in the planar limit. Based on the strong coupling asymptotic BA first proposed in \cite{afs}, and using earlier results \cite{ptt, bt, hl, bhl, es}, the all loop ABA sector was proposed in \cite{bes}. The main ingredient
in the ABA is the all-loop dressing phase which interpolates between strong and weak coupling. A rigorous proof of the relationship between strong/weak coupling expansions of the coefficients $c_{r,s}$ in the dressing phase was given in \cite{kt}. So far the ABA proposal passed all tests both at strong and weak coupling.

Much of the progress in testing the all-loop ABA relates the twist two gauge operators of the type $tr(\Phi D_{+}^S \Phi)$ to the folded string solution \cite{gkp}. More specifically, detailed tests were performed in the large limit of the semi-classical spin $\mathcal{S}$, in which string theory and ABA computations match to two loops in strong coupling expansion \cite{ft1, bftt, rtt,bbks,bkk}. The checks to two loops were successfully extended to include the spin $J$ in $S^1 \subset S^5$ in \cite{ftt, ck, g1, grrtv}.  Beyond the long string limit the string solution is no longer homogenous, which makes even the $1$-loop string computation difficult. Recently, the exact $1$-loop correction to the energy for the folded spinning string was obtained \cite{bdfpt}. It is of interest to check explicitly that computation against the corresponding $1$-loop result from ABA for arbitrary spins $\mathcal{S}, \mathcal{J}$. In this paper we recover from ABA at leading order in strong coupling the exact folded string solution valid for any semi-classical spins   $\mathcal{S}=\frac{S}{\sqrt{\lambda}}, \mathcal{J}=\frac{J}{\sqrt{\lambda}}$ including $\mathcal{J}=0$. Going to $1$-loop correction, one expects matching based on the discussion in \cite{gv}.
To explicitly and fully match the $1$-loop correction from string theory \cite{bdfpt} one needs to consider the corrected asymptotic BA equations at $1$-loop proposed recently in \cite{g}.  One particular interesting case from the BA side is to find the $1$-loop and higher loop corrections to the energy of the folded string in the short string limit (small $\mathcal{S}$) with arbitrary $\mathcal{J}$, and match it to the known results \cite{tt, g2, bts, bdfpt}.

Beyond twist two operators, the $SL(2)$ sector contains operators of higher twist which have also been investigated. For example operators of the type
\beq\la{opa}
\cO = \tr\left( D_+^{\frac{S}{n}}\Phi\ D_+^{\frac{S}{n}}\Phi\ldots D_+^{\frac{S}{n}}\Phi\  \right)
\eeq
are described,
on the string side, by the spiky string solutions \cite{k}. There has been some recent progress in this area. In \cite{Jevicki}
the solutions
were shown to correspond to multi-soliton solutions of a generalized sinh-Gordon model. This allowed the construction of new, more general, solutions
where the spikes move with respect to each other opening up the possibility to study a whole new class of operators. The relation with field theory was examined
in detail in \cite{Dorey1} where the elliptic curves associated with the classical solution where analyzed and a map was proposed
to a similar structure emerging from the study of the field theory operators.

In \cite{fkt} the spiky string solution was analyzed in detail by using the all loop ABA at leading and subleading orders in large $S$ expansion. In this limit the $1$-cut solution of the ABA equations was found, and matching with the string theory result was shown. Furthermore in \cite{fkt} the spiky string $2$-cut solution at $1$-loop weak coupling Bethe ansatz was discussed. Unlike the folded string, the spiky string requires asymmetric $2$-cut solutions, which in turn give asymmetric distributions of the Bethe roots. One of the main ingredients in the ABA are the integers $n_k$ that appear after taking the logarithm of the Bethe ansatz equations. It was proposed in \cite{fkt} that the $n_k$ corresponding to the left and right cuts are in fact related to the left/right bosonic mode numbers of the corresponding string solution in the flat space limit, i.e. near the center of the $AdS$. Following the general procedure of solving integral BA equation used in \cite{fkt}, in this paper we check this proposal directly at strong coupling by finding explicitly the $2$-cut ABA solution, and matching it to the known spiky string solution. We found a system of equations that can in principle be solved for the four ends of the cuts $[d,c], [b,a]$, and thus the energy can be obtained. For $\mathcal{J}=0$ the cuts extend as $[d,-1]$ and $[1,a]$, and we find the relationship between $a,d$ and the maximum ($\rho_1$) and minimum ($\rho_0$) radial extensions of the string $a=e^{2 \rho_1- 2 \rho_0}$,
$-d=e^{2 \rho_1+ 2 \rho_0}$. It would be interesting to derive the classical string equations from the ABA integral equation perhaps using the coherent state approach as in \cite{kru}. This would allow to see how the shape of the string is encoded in the ABA approach.

A general ansatz for finding rigid string solutions in $AdS_3 \times S^1$ was used in \cite{iktt}. Spiky string solutions characterized by the spins ($S,J$) in $AdS_3$ and $S^1 \subset S^5$, winding number $m$ in $S^1$, and number of spikes $n$ were found whose energy can be written as
\begin{equation}
E= S+J + \gamma (S,J, m,n; \lambda)
\end{equation}
On the asymptotic Bethe ansatz side, the $1$-cut solution with non-trivial winding was discussed in \cite{szz} where the rational $(S,J)$ solution was found. In this paper, using the ABA equations, we find the spiky $2$-cut solutions with arbitrary winding $m$, spikes $n$, and angular momenta $S$ and $J$. To check the matching with solutions found in \cite{iktt} we consider the scaling limit
\begin{equation}
\mathcal{S}, \mathcal{J},m \rightarrow \infty \ \ \ \ \texttt{with} \ \ \ \ \frac{\mathcal{J}}{\ln \mathcal{S}}=\texttt{fixed}, \ \ \ \ \ \frac{m}{\ln \mathcal{S}}=\texttt{fixed}
\end{equation}
Taking this limit in both solutions from string sigma model and ABA we find perfect matching of the energy
\begin{equation}
\mathcal{E}-\mathcal{S}=\frac{n}{2 \pi} \sqrt{1+ \frac{4 \pi^2}{n^2}\bigg(\frac{\mathcal{J}^2}{\ln^2 \mathcal{S}}+\frac{m^2}{\ln^2 \mathcal{S}}\bigg)}\ln \mathcal{S}+...  \label{scal3}
\end{equation}
This limit corresponds to the string touching the boundary, or from the ABA side meaning cuts very long, i.e. $a, -d \rightarrow \infty$.
For $n=2$ the result (\ref{scal3}) corresponds to the folded string with winding. Such a scaling limit was found very recently in
\cite{grrtv} where the asymptotic open string solution was considered.

This result further checks that the strong coupling ABA of \cite{afs} indeed captures such more sophisticated solutions. With appropriate cuts it also captures the $\mathcal{J}=0$ solutions. An interesting open problem is to see whether and/or how one can use this information to obtain an asymptotic Bethe ansatz to describe all classical string solutions in $AdS_3$ with $\mathcal{J}=0$. In particular it would be nice to find the solutions in \cite{Jevicki} within the BA approach.

The rest of the paper is organized as follows. In section 2 we review the ABA equations at leading order in strong coupling, and present the general formulas that describe two-cut solutions. In section 3 we solve the ABA equations for the symmetric two-cut solution, and match the solution to the known folded spinning string result. Following the
simpler example discussed in section 3, we find the spiky string solution with arbitrary spins $\mathcal{S}, \mathcal{J}$, and show the matching with the know spiky string solution. In section 5 we extend the analysis to solutions that include also a winding in $S^1 \subset S^5$. We find from ABA a system of equations which describe such solutions. Furthermore, we consider the scaling limit with $\mathcal{S}, \mathcal{J}, m$ large and show the matching to string results in this limit. In Appendix A we present a proof of some particular relations among elliptic integrals that were used in the main text.

\renewcommand{\theequation}{2.\arabic{equation}}
 \setcounter{equation}{0}

\setcounter{equation}{0} \setcounter{footnote}{0}
\setcounter{section}{1}

\section{Two-cut solutions in asymptotic Bethe Ansatz at leading order in strong coupling}

The Bethe equations in $SL(2)$ sector are  ($g^2=\frac{\lambda}{8 \pi^2}$)  \cite{afs, es, bes}
\begin{eqnarray}
\bigg(\frac{x_k^{+}}{x_k^{-}}\bigg)^J= \prod_{j\neq k}^S \bigg(\frac{x_k^{-}-x_j^{+}}{x_k^{+}-x_j^{-}}\bigg) \frac{1- \frac{g^2}{2 x_k^{+} x_j^{-}}}{1-\frac{g^2}{2 x_j^{+} x_k^{-}}}e^{2 i \theta(x_k,x_j)}  \label{be}
\end{eqnarray}
with the phase
\begin{equation}
\theta(x_k,x_j)=\sum_{r=2}^{\infty} \sum_{s=r+1}^{\infty} \bigg(\frac{g^2}{2}\bigg)^{(r+s-1)/2}c_{r,s}(g)[q_r(x_k) q_s(x_j)-q_r(x_j)q_s (x_k)]
\end{equation}
where the charges $q_r$ are
\begin{equation}
q_r(x_k)= \frac{i}{r-1}\bigg(\frac{1}{(x_k^{+})^{r-1}}-\frac{1}{(x_k^{-})^{r-1}}\bigg), \quad \quad Q_r= \sum_k q_r(x_k)
\end{equation}
The coefficients $c_{r,s}(q)$ are expanded at strong coupling as
\begin{equation}
c_{r,s}(\lambda)= c_{r,s}^{(0)} + \frac{1}{\sqrt{\lambda}}c_{r,s}^{(1)}+...
\end{equation}
where
\begin{equation}
c_{r,s}^{(0)}=\delta_{r,s+1}, \quad \quad c_{r,s}^{(1)}= -4 [1- (-1)^{r+s}]\frac{(r-1)(s-1)}{(s+r-2)(s-r)}
\end{equation}
The momentum condition is
\begin{equation}
\prod_{k=1}^S \bigg(\frac{x_k^{+}}{x_k^{-}}\bigg)=1  \label{p}
\end{equation}
while the string energy is
\begin{equation}
E-S-J = \frac{\lambda}{8 \pi^2}Q_2   \label{en}
\end{equation}

The relationship between $x$ and the rapidity variable $u$ is
\begin{equation}
x^{\pm}= x(u \pm \frac{i}{2}), \quad \quad x(u)=\frac{u}{2}+\frac{u}{2}\sqrt{1- \frac{2 g^2}{u^2}},\quad \quad u(x)=x+ \frac{g^2}{2 x}
\end{equation}

We are interested in the leading order at strong coupling, $S \sim \sqrt{\lambda}$, $J \sim \sqrt{\lambda}$. In this case the Bethe roots $u_k$ scale as $\sqrt{\lambda}$. We keep only the leading order $c_{r,s}^{(0)}$ in what follows.
As we are interested in large $g$, we perform the following rescaling  \cite{g1}
\begin{equation}
x \rightarrow \frac{g}{\sqrt{2}}x
\end{equation}
After this rescaling $x_k \sim 1$
\begin{equation}
x_k= 2 \pi \frac{u_k}{\sqrt{\lambda}}+\sqrt{4 \pi^2 \bigg(\frac{u_k}{\sqrt{\lambda}}\bigg)^2-1}, \quad \quad u_k= \frac{g (x_k^2+1)}{\sqrt{2} x_k} \label{aj}
\end{equation}
Using the last expression for $u_k$ and expanding in large $g$ we get
\begin{equation}
x_k^{\pm}= x_k \pm \frac{i x_k^2}{\sqrt{2} (x_k^2-1)}\frac{1}{g}+\frac{x_k^3}{2 (x_k^2-1)^3}\frac{1}{g^2} +...
\end{equation}

Using these expressions in  the string Bethe equations, we obtain the following equations at the leading order in large $g$
\begin{equation}
\frac{2 \sqrt{2}}{g}\sum_{j \neq k}^S\frac{1-\frac{1}{x_j x_k}}{(x_k-x_j) (1- \frac{1}{x_k^2})(1-\frac{1}{x_j^2})} = 2 \pi n_k - \frac{\sqrt{2} J x_k}{x_k^2-1} \frac{1}{g}  \label{alp}
\end{equation}

Since in the strong coupling limit $S$ is large, i.e. there is a large number of Bethe roots, we introduce a density distribution in $x$ as $\rho(x)= \frac{\sqrt{2}}{g}\sum_k \delta(x-x_k)$. Therefore equation (\ref{alp}) becomes
\begin{equation}
 \int d x' \rho(x') \frac{1-\frac{1}{x x'}}{(x-x')(1-\frac{1}{x'^2})}=  \pi n(x) \big(1- \frac{1}{x^2}\big)- 2 \pi \frac{J}{\sqrt{\lambda}}\frac{1}{x} \label{qoi}
\end{equation}
On the right hand side we observe the appearance of the semi-classical spin $\mathcal{J}=\frac{J}{\sqrt{\lambda}}$. While a $2$-cut solution cannot be obtained by setting $\mathcal{J}=0$ (see below), we can still obtain the folded and spiky string solution with $\mathcal{J}=0$ by taking the ends of the cuts at $\pm 1$.

To see the domain for $x$ we rescale $u_k$ in (\ref{aj}) as $u_k \rightarrow \sqrt{2} g u_k$, so that now $u_k \sim 1$. Then we have
\begin{equation}
x_k= u_k + \sqrt{u_k^2-1}
\end{equation}
While $u_k$ are always real, we observe that when $u_k^2 \geq 1$, $x_k$ is also real, while if $u_k \leq 1$, then $x_k$ is complex
$x_k= u_k + i \sqrt{1-u_k^2}= e^{i \alpha}$, $\tan \alpha= \frac{\sqrt{1-u_k^2}}{u_k}$.
To find the $1$-cut and $2$-cuts spiky string solutions in this limit using (\ref{qoi}) we need to choose the right cuts along the $x$ variable. The $x$ variable goes along the real line except near origin where it goes on a half unit circle. Throughout this paper we take the cuts outside the interval $[-1,1]$.

Returning to equation (\ref{qoi})  it is convenient to introduce
\begin{equation}
\tilde{\rho}(x)=\frac{x^2}{x^2-1}\rho(x), \quad \quad \quad  \tilde{n}(x)=n(x)-2 \mathcal{J}\frac{x}{x^2-1}, \quad \quad \quad \mathcal{J}=\frac{J}{\sqrt{\lambda}}
\end{equation}
\begin{equation}
\int d x' \frac{\tilde{\rho}(x')}{x-x'}(1-\frac{1}{x x'})=\pi (1-\frac{1}{x^2})\tilde{n}(x)  \label{bas}
\end{equation}

The classical string energy (\ref{en}) can be written as
\begin{equation}
E-S-J = \frac{\sqrt{\lambda}}{2 \pi} \int dx \frac{\rho(x)}{x^2-1}=\frac{\sqrt{\lambda}}{2 \pi} \int dx \frac{\tilde{\rho}(x)}{x^2}
\end{equation}
The momentum condition (\ref{p}) lead to
\begin{equation}
\int d x \rho(x) \frac{x}{x^2-1}=\int d x \frac{\tilde{\rho}(x)}{x}=0  \label{pap}
\end{equation}
Finally the normalization condition reads
\begin{equation}
\int d x \tilde{\rho} (x) (1-\frac{1}{x^2})= 4 \pi \mathcal{S}, \quad \quad \quad \mathcal{S}=\frac{S}{\sqrt{\lambda}}
\end{equation}
Let us notice that the resulting equations have indeed the semi-classical form as expected. More precisely, a solution for $\tilde{\rho}$ only depends on semi-classical parameters $\mathcal{J},\mathcal{S}$, and then the energy can be expressed as $\mathcal{E}=\mathcal{E}(\mathcal{J},\mathcal{S})$. The unknown unphysical parameters entering the equations are only the positions of the cuts which
are to be solved for. The essential input in the BA equations is the mode number $n(x)$.
Let us observe that upon using the momentum condition (\ref{pap}), the resulting integral equation (\ref{bas}) is precisely the same as the equation found in \cite{kz}.

\bigskip

\begin{figure}
\begin{center}
\epsfig{file=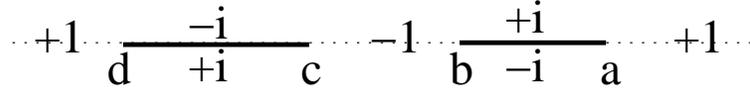, width=10cm}
\end{center}
\caption{Phase of the function $F(w)$ defined in the text, when $w$ approaches the real axes. Along the two cuts, the imaginary part
changes sign as depicted. }
\label{fig:cuts1}
\end{figure}

To solve (\ref{bas}) we employ the method described in \cite{fkt}. While below we specialize to two cuts, let us mention that the method can be used for any number of cuts.
To be specific lets consider the two cuts
$C_1=[d,c]$, $C_2=[b,a]$ with $d<c<b<a$ and the following function:
\beq
 F(w) = \sqrt{w-a}\sqrt{w-b}\sqrt{w-c}\sqrt{w-d}
\eeq
where the square roots are defined with a standard cut on the negative real axis. In figure \ref{fig:cuts1} we show the phase of $F(w)$ for
$w$ close to the real axis. As we mentioned already, here we take the cuts outside the interval $[-1,1]$, namely we take $b >1$, $c<-1$.

\begin{figure}
\begin{center}
\epsfig{file=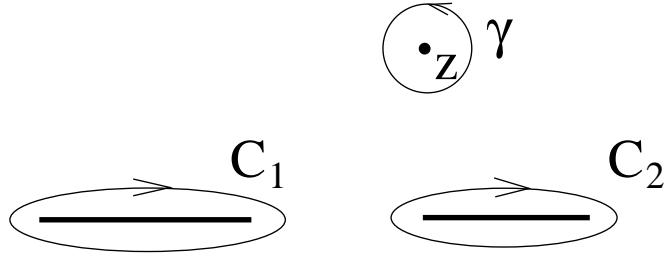, width=10cm}
\end{center}
\caption{Contours of integration surrounding $z$ and the two cuts.
 }
\label{fig:cuts2}
\end{figure}
Now we can write
\beq
 \Gg(w) = \frac{1}{\pi} \int_{C_1\cup C_2} \frac{F(w)}{F(x+i\epsilon)} \frac{\tilde{n}(x)}{x-w} dx  \label{resolvent}
\eeq
It is easy to verify that $\Gg(w)$ so defined has cuts in $C_1$ and $C_2$ and satisfies the equation
\beq
 \Gg(x\pm i\epsilon) = \pm \tilde{\rho}(x) + i \tilde{n}(x), \ \ \ \ x\in C_1\cup C_1 , \ \ \ \epsilon\rightarrow 0^+.
\label{jumps1}
\eeq
with
\beq
\tilde{\rho}(x) = \frac{1}{\pi} \ppint \left|\frac{F(x)}{F(x')}\right|\frac{\tilde{n}(x')}{x'-x} dx'
\label{rhosol}
\eeq

Now consider an arbitrary point $z$ (away from the cuts) and a small contour
$\gamma$ encircling it as shown in figure \ref{fig:cuts2}. We have
\beq
(1-\frac{1}{z^2}) \Gg(z) = \frac{1}{2\pi i} \oint_\gamma \frac{\Gg(w)}{w-z} (1-\frac{1}{w z}) dw
\eeq
Deforming the contour we obtain
\begin{eqnarray}
(1-\frac{1}{z^2}) \Gg(z) &=& -\res{\frac{\Gg(w)}{w-z}(1-\frac{1}{w z})}{w=\infty} -\res{\frac{\Gg(w)}{w-z}(1-\frac{1}{w z})}{w=0}\nonumber\\
&+& \frac{1}{2\pi i} \oint_{C_1\cup C_2} \frac{\Gg(w)}{w-z} (1-\frac{1}{w z}) dw
\end{eqnarray}
From (\ref{jumps1}) we further have
\begin{eqnarray}
(1-\frac{1}{z^2}) \Gg(z) &=& -\res{\frac{\Gg(w)}{w-z}(1-\frac{1}{w z})}{w=\infty} -\res{\frac{\Gg(w)}{w-z}(1-\frac{1}{w z})}{w=0}\nonumber\\
&+& \frac{1}{\pi i} \int_{C_1\cup C_2} \frac{\tilde{\rho}(x')}{x'-z} (1-\frac{1}{x' z}) dw
\end{eqnarray}
From this result and (\ref{jumps1}) we find that, if $z$ approaches one of the cuts from above ($z=x+i \epsilon$), then:
\beqa
(1-\frac{1}{x^2})\tilde{\rho}(x) &+& i (1-\frac{1}{x^2})\tilde{n}(x) =(1-\frac{1}{x^2}) \Gg(x+i\epsilon) = (1-\frac{1}{x^2})\tilde{\rho}(x)\nonumber\\
&-&\res{\frac{\Gg(w)}{w-z}(1-\frac{1}{w z})}{w=\infty}
                      - \res{\frac{\Gg(w)}{w-z}(1-\frac{1}{w z})}{w=0}\nonumber\\
                       &+& \frac{1}{\pi i} \ppint_{C_1\cup C_2} \frac{\tilde{\rho}(x')}{x'-x}(1-\frac{1}{x x'}) dx',
                       \ \ \ \epsilon\rightarrow 0^+
\eeqa
So we obtain that $\tilde{\rho}(x)$ solves the problem (\ref{bas}) under the conditions ($z=x+i \epsilon$)
\beq
\res{\frac{\Gg(w)}{w-z}(1-\frac{1}{w z})}{w=\infty} = 0, \quad \quad \res{\frac{\Gg(w)}{w-z}(1-\frac{1}{w z})}{w=0}  \label{fii}
\eeq
Let us note that this analysis holds for any function $\tilde{n}(x)$.
The residue at zero can be computed by expanding at small $w$
\beq
 \Gg(w) = G(0) + H_1 w  +\ldots
\eeq
The second residue in (\ref{fii}) is simply
\begin{equation}
\res{\frac{\Gg(w)}{w-z}(1-\frac{1}{w z})}{w=0}= \frac{G(0)}{x^2}  \label{reszero}
\end{equation}
The residue at infinity can be computed by expanding at large $w$
\beq
 \Gg(w) = \Gg_0 w+ \Gg_1 + \frac{1}{w} \Gg_2 +\ldots
\eeq
so that the first residue in (\ref{fii}) is
\begin{equation}
\res{\frac{\Gg(w)}{w-z}(1-\frac{1}{w z})}{w=\infty}=-x G_0 -G_1 + \frac{G_0}{x}
\end{equation}
Thus to have solution for $\tilde{\rho}$ given by (\ref{rhosol}) we require (according to (\ref{fii}))
\begin{equation}
G_0=G_1=G(0)=0
\end{equation}
The normalization and the momentum conditions can be written as
\begin{equation}
2 \pi \mathcal{S}=\frac{1}{2}\int d x \tilde{\rho} (x) (1-\frac{1}{x^2})=\frac{1}{2}(i \pi G_0 -i \pi G_2 -i \pi H_1)  \label{norm}
\end{equation}
\begin{equation}
\int d x \frac{\tilde{\rho}(x)}{x}= i \pi G(0)-i \pi G_1=0  \label{mom}
\end{equation}
Finally, the energy can be written as
\begin{equation}
(\mathcal{E}-\mathcal{S}-\mathcal{J})2 \pi=\int dx \frac{\tilde{\rho}(x)}{x^2}=-i \pi G_0 + i \pi H_1
\end{equation}
Thus, we obtain a set of equations which need to be solved for the unknown parameters $a,b,c,d$. The only input that we need to specify for a given particular solution is the function $n(x)$.

\section{Folded spinning string solution}

\renewcommand{\theequation}{3.\arabic{equation}}
 \setcounter{equation}{0}

As a simple example of the method described in the previous section, here we solve the ABA equation with two symmetric cuts, which corresponds to the spinning folded string solution in the $SL(2)$ sector. The relevant equations were found in \cite{ck} by using a certain scaling of the distribution of Bethe roots; by using a different scaling here we obtain explicitly the $\mathcal{J}=0$ folded string solution as well.
For the folded string we set $d=-a$, $c=-b$. It is convenient to define the function
\begin{equation}
|F(x)|=\sqrt{|(x^2-a^2)(x^2-b^2)|}
\end{equation}
Since in the flat space the folded string solution has equal left and right wave numbers we take
\begin{eqnarray}
n(x)= \left\{
        \begin{array}{ll}
          -1, & -a < x <-b \\
          1, & b< x< a
        \end{array}
      \right.
\end{eqnarray}

The following integrals\footnote{In this paper we use the following definitions of the elliptic integrals
    $K[\mu]=\int_0^{\frac{\pi}{2}} \frac{d \alpha}{\sqrt{1-\mu \sin^2 \alpha}}$, $E[\mu]=\int_0^{\frac{\pi}{2}} d \alpha \sqrt{1-\mu \sin^2 \alpha}$, $\Pi[\nu,\mu]=\int_0^{\frac{\pi}{2}} \frac{d \alpha}{(1- \nu \sin^2
 \alpha)\sqrt{1-\mu \sin^2 \alpha}}$.} are of use
\begin{equation}
A_1= \int_{-a}^{-b} \frac{d x}{|F(x)|}=\int_b^a \frac{d x}{|F(x)|}=\frac{1}{a}K[q], \quad \quad
A_2= \int_{-a}^{-b} \frac{x dx}{|F(x)|}=-\int_{b}^{a} \frac{x dx}{|F(x)|}=-\frac{\pi}{2}
\end{equation}
\begin{equation}
A_3= \int_{-a}^{-b} \frac{dx}{x |F(x)|}=- \int_{b}^{a} \frac{dx}{x |F(x)|}=-\frac{\pi}{2 a b}, \quad \quad
A_4= \int_{-a}^{-b} \frac{x^2 dx }{ |F(x)|}=\int_b^a \frac{x^2 dx}{ |F(x)|}=a E[q]
\end{equation}
\begin{equation}
A_5= \int_{-a}^{-b} \frac{dx}{x^2 |F(x)|}=\int_b^a \frac{dx }{x^2 |F(x)|}=\frac{1}{a b^2}E[q]
\end{equation}
\begin{equation}
A_6= \int_{-a}^{-b} \frac{dx x}{(x^2-1) |F(x)|}=-\int_b^a \frac{dx x}{(x^2-1) |F(x)|}=-\frac{\pi}{2 \sqrt{(a^2-1)(b^2-1)}}
\end{equation}
\begin{equation}
A_7= \int_{-a}^{-b} \frac{dx }{x(x^2-1) |F(x)|}=- \int_b^a \frac{dx }{x(x^2-1) |F(x)|}=-\frac{\pi}{2}\bigg(\frac{1}{\sqrt{(a^2-1)(b^2-1)}}-\frac{1}{ab}\bigg)
\end{equation}
\begin{equation}
A_8= \int_{-a}^{-b} \frac{dx x^3}{(x^2-1) |F(x)|}=- \int_b^a \frac{dx x^3}{(x^2-1) |F(x)|}=-\frac{\pi}{2}\bigg(\frac{1}{\sqrt{(a^2-1)(b^2-1)}}+1\bigg)
\end{equation}
where $q=\frac{a^2-b^2}{a^2}$.
Let us compute the relevant parameters that enter the residues. We find
\begin{equation}
G(0)=G_1=0, \quad \quad \quad G_0=-\frac{1}{i \pi}(2 A_1 + 4 \mathcal{J}A_6)
\end{equation}
In order to have a solution we need $G_0=0$, which gives the condition
\begin{equation}
K[q]=\frac{a \pi}{\sqrt{(a^2-1)(b^2-1)}}\mathcal{J}  \label{spinJ}
\end{equation}

The momentum equation becomes
\begin{equation}
\int d x \frac{\tilde{\rho}(x)}{x}= i \pi G(0)-i \pi G_1=0
\end{equation}
which is satisfied automatically for a solution $\tilde{\rho}(x)$.

For the energy we obtain
\begin{eqnarray}
(\mathcal{E}-\mathcal{S}-\mathcal{J})2 \pi&=&\int dx \frac{\tilde{\rho}(x)}{x^2}=-i \pi G_0 + i \pi H_1=-a b (2 A_5+4 \mathcal{J} A_7)\nonumber\\
&=&
2 \pi \mathcal{J} a b \bigg(\frac{1}{\sqrt{(a^2-1)(b^2-1)}}-\frac{1}{a b}\bigg)-\frac{2}{b}E[q]  \label{ener}
\end{eqnarray}
where we used $H_1$ defined as
\begin{equation}
H_1= -\frac{1}{\pi}\int d x \frac{\tilde{n}(x)}{F(x+i \epsilon)}\frac{a b}{x^2}=-\frac{ab}{i \pi}(2 A_5 + 4 \mathcal{J} A_7)
\end{equation}
The spin can be written as
\begin{equation}
2 \pi \mathcal{S}=\frac{1}{2}(i \pi G_0 -i \pi G_2 -i \pi H_1)=A_4+ a b A_5 + 2 \mathcal{J}(a b A_7 +A_8)=  (a+\frac{1}{b})E[q]-\pi \mathcal{J}\frac{(1+a b)}{\sqrt{(a^2-1)(b^2-1)}}  \label{spinS}
\end{equation}

Therefore we obtain the semiclassical equations (\ref{spinJ}), (\ref{spinS}), (\ref{ener}). One needs to solve  (\ref{spinJ}), (\ref{spinS}) for $a,b$ and then plug the result in (\ref{ener}) to obtain $\mathcal{E}=\mathcal{E}(\mathcal{S},\mathcal{J})$.

The particular case $\mathcal{J}=0$ can be obtained in the limit $b \rightarrow 1$ assuming $a>b>1$\footnote{If $b<a<1$ we can take $a\rightarrow 1$ to have $\mathcal{J}=0$. However, it turns out that the energy in this case is negative, so we discard such solution.}. We observe that this limit is indeed possible in  (\ref{spinJ}), therefore the motion only in $AdS$ can still be described\footnote{Note that starting directly with (\ref{qoi}) and setting $\mathcal{J}=0$ one cannot obtain a two-cut solution.} using the SL(2) ABA equations with cuts that start at $x=1$.
In this limit we obtain the energy and spin
\begin{equation}
2 \pi (\mathcal{E}-\mathcal{S})= 2 \bigg(K[1-\frac{1}{a^2}]-E[1-\frac{1}{a^2}] \bigg), \quad \quad
2 \pi \mathcal{S}= (a+1) E[1-\frac{1}{a^2}]-(1+\frac{1}{a}) K[1-\frac{1}{a^2}]
\end{equation}
In the limit $a \rightarrow \infty $ these expressions precisely reproduce the long string limit of the folded string solution, while when $a \rightarrow 1$ we obtain the short string limit. To see the precise folded string form of equations we consider the elliptic integral modular transform relations\footnote{Precisely the same modular transformation relations were used for the folded string at $1$-loop in weak coupling in \cite{bfst}.}
\begin{equation}
K[-\frac{1}{\eta}]=(1-q)^{1/4} K[q], \quad  E[-\frac{1}{\eta}]=\frac{1}{2}(1-q)^{-1/4} E[q]+\frac{1}{2}(1-q)^{1/4} K[q], \quad  \frac{1}{\eta}=\frac{(1-\sqrt{1-q})^2}{4 \sqrt{1-q}}  \label{mod}
\end{equation}
For $q=1-\frac{1}{a^2}$ we obtain the folded string solution equations \cite{bftt, bfst}
\begin{equation}
\mathcal{E}-\mathcal{S}=\frac{2}{\pi}\sqrt{\frac{\eta+1}{\eta}}\bigg[E[-\frac{1}{\eta}](\frac{1}{\sqrt{1+\eta}}-1)+K[-\frac{1}{\eta}]\bigg], \quad
\mathcal{S}=\frac{2}{\pi}\sqrt{\frac{\eta+1}{\eta}}\bigg[E[-\frac{1}{\eta}]-K[-\frac{1}{\eta}]\bigg]
\end{equation}
The small $\eta$ limit corresponds to long string, while the large $\eta$ to short string limit. $\eta$ is related to the maximal extension in the radial coordinate in $AdS$, as $\eta \sinh^2 \rho_1=1$. The relationship between the Bethe ansatz parameter $a$ and $\rho_1$ is $a=e^{2 \rho_1}$.

To analyze the $\mathcal{J}\neq 0 $ solution let us now return to the equations (\ref{spinJ}), (\ref{spinS}), (\ref{ener}) and rewrite the last two ones as
\begin{equation}
2 \pi \mathcal{E}= (b-\frac{1}{a})K[q]+(1-\frac{1}{b})E[q], \quad \quad 2 \pi \mathcal{S}=(a+\frac{1}{b})E[q]-(b+\frac{1}{a})K[q]
\end{equation}
One can check numerically that solutions of equations (\ref{spinJ}), (\ref{spinS}), (\ref{ener}) indeed exist for $1< b<a$. For $b<a<1$ the energy is negative, thus this case is not physical.

Using again the modular transformations (\ref{mod}) for $q=1-\frac{b^2}{a^2}$ we obtain precisely the folded string
solution \cite{bfst}
\begin{equation}
\frac{\mathcal{E}^2}{E^2[-\frac{1}{\eta}]}-\frac{\mathcal{J}^2}{K^2[-\frac{1}{\eta}]}=\frac{4}{\pi^2}\frac{1}{\eta}, \quad \quad \frac{\mathcal{S}^2}{(K[-\frac{1}{\eta}]-E[-\frac{1}{\eta}])^2}-\frac{\mathcal{J}^2}{K^2[-\frac{1}{\eta}]}=
\frac{4}{\pi^2}\frac{\eta+1}{\eta}
\end{equation}
The ABA parameters $a$, $b$ are related to string maximal extension, $\rho_1$ (the folded string extends from $\rho=0$ to $\rho=\rho_1$), as $\frac{a}{b}=e^{2 \rho_1}$.

\renewcommand{\theequation}{4.\arabic{equation}}
 \setcounter{equation}{0}

\section{Spiky string solution}

To obtain spiky strings we take two cuts with different lengths $[d,c]$ and $[b,a]$. Again we take the cuts outside the $[-1,1]$ interval, i.e. $1<b<a$ and $d<c<-1$.
As proposed in \cite{fkt} we take the function $n(x)$ to be the left and right wave numbers of the spiky string solution in flat space \cite{k}
\begin{eqnarray}
n(x)= \left\{
        \begin{array}{ll}
          -1, & d < x <c \\
          n-1, & b< x< a
        \end{array}
      \right.
\end{eqnarray}
where $n$ is the number of spikes.
Starting with (\ref{resolvent}) we can compute $G(w)$ by performing the integral and obtain
\begin{eqnarray}
G(w)&=&- \frac{2 i n}{\pi \sqrt{(a-c)(b-d)}}\frac{F(w)}{(b-w)(w-c)}\bigg((b-c)\Pi[\frac{(a-b)(w-c)}{(a-c)(w-b)},r]+(w-b)K[r]\bigg)\nonumber\\
&-&i -2 i \mathcal{J}\bigg[\frac{w}{w^2-1}+\frac{F(w)}{2}\bigg(\frac{1}{F(1)(1-w)}-\frac{1}{F(-1)(1+w)}\bigg)\bigg]
\end{eqnarray}
where $r=\frac{(a-b)(c-d)}{(a-c)(b-d)}$ and $F(0)=-\sqrt{a b c d}$
\begin{equation}
F(1)=-\sqrt{(1-a)(1-b)(1-c)(1-d)}, \quad F(-1)=-\sqrt{(1+a)(1+b)(1+c)(1+d)}
\end{equation}
We can now expand for small and large $w$ and obtain $G(0), G_0, G_1, G_2, H_1$
\begin{equation}
G_0=  \frac{2 i n}{\pi \sqrt{(a-c)(b-d)}}K[r] +i \mathcal{J}\bigg(\frac{1}{F(1)}+\frac{1}{F(-1)}\bigg)
\end{equation}
\begin{eqnarray}
G_1&=& -i  + \frac{2 i n}{\pi} \frac{b-c}{\sqrt{(a-c)(b-d)}}\Pi + \frac{2 i n}{\pi} \frac{c}{\sqrt{(a-c)(b-d)}} K[r] + i \mathcal{J}\bigg(\frac{1}{F(1)}-\frac{1}{F(-1)}\bigg)
\end{eqnarray}
\begin{eqnarray}
G_2&=& -2 i \mathcal{J}+\frac{2 i n}{\pi}\frac{(b^2-c^2)}{\sqrt{(a-c)(b-d)}}\Pi + \frac{2 i n}{\pi}\frac{(b-c)^2 (a-b)}{(a-c)\sqrt{(a-c)(b-d)}}\Pi'\nonumber\\
&+& \frac{2 i n}{\pi}\frac{c^2}{\sqrt{(a-c)(b-d)}}K[r]+ i \mathcal{J}\bigg(\frac{1}{F(1)}+\frac{1}{F(-1)}\bigg)-\frac{i}{2}(a+b+c+d)
\end{eqnarray}

The $w \rightarrow 0$ expansions give
\begin{eqnarray}
G(0)&=&-i + F(0)\bigg[\frac{2 i n}{\pi}\frac{b-c}{b c \sqrt{(a-c)(b-d)}}\bar{\Pi} + \frac{i \mathcal{J}}{c}\bigg(\frac{1-c}{F(1)}+\frac{1+c}{F(-1)}\bigg)\bigg]
\end{eqnarray}
\begin{eqnarray}
H_1&=& 2 i \mathcal{J}+\frac{i}{2}\bigg(\frac{1}{b}+\frac{1}{c}-\frac{1}{a}-\frac{1}{d}\bigg)- \frac{2 i n}{\pi}\frac{(b-c)^2 (a-b) c}{b^2 c^2 (a-c) b \sqrt{(a-c)(b-d)}}\bar{\Pi}' F(0)\nonumber\\
&-& \frac{i \mathcal{J}}{b c}F(0) \bigg(\frac{(1-c)(1-b)}{F(1)}+\frac{(1+c)(1+b)}{F(-1)}\bigg)
\end{eqnarray}
where
\begin{equation}
\Pi=\Pi[\frac{a-b}{a-c},r], \quad \quad \bar{\Pi}[\frac{(a-b)c}{(a-c)b},r]
\end{equation}
and prime denotes the derivatives of elliptic $\Pi[n,m]$ in respect to $n$
\begin{equation}
\Pi'= \frac{\partial \Pi[n,m]}{\partial n}=\frac{1}{2 (m-n)(n-1)}\bigg[E[m] + \frac{(m-n)}{n}K[m]+\frac{n^2-m}{n}\Pi[n,m]\bigg]  \label{deriv}
\end{equation}

Equations $G_0=G(0)=G_1=0$ along with the normalization condition (\ref{norm})
are the equations necessary to find the unknown parameters $a,b,c,d=a,b,c,d(n,\mathcal{J},\mathcal{S})$. Once found $a,b,c,d$ can be used in the semi-classical expression for the energy to find
$\mathcal{E}=\mathcal{E}(\mathcal{J},\mathcal{S},n)$.

Let us investigate in detail the particular case with $\mathcal{J}=0$. As in the case of folded string, to take $\mathcal{J}\rightarrow 0$ we take $b \rightarrow 1$, $c \rightarrow -1$, so $r=\frac{(a-1)(d+1)}{(a+1)(d-1)}$.
Taking this limit we obtain the following equations
\begin{equation}
\frac{4 n}{\pi}\frac{\Pi-\bar{\Pi}}{\sqrt{(a+1)(1-d)}}=1- \frac{1}{\sqrt{-a d}}  \label{spikes}
\end{equation}
\begin{equation}
\Delta=\mathcal{E}-\mathcal{S}=\frac{i}{2}H_1=\frac{1}{4}(\frac{1}{a}+\frac{1}{d})+\frac{4 n}{\pi}\frac{a-1}{a+1}\frac{\sqrt{-a d}}{\sqrt{(a+1)(1-d)}}\bar{\Pi}'  \label{ena}
\end{equation}
\begin{eqnarray}
\mathcal{S}=-\frac{i}{4}(G_2+H_1)&=& -\frac{1}{8}(a+d)-\frac{1}{8}(\frac{1}{a}+\frac{1}{d})+\frac{2 n}{\pi}\frac{a-1}{a+1}\frac{\Pi'}{\sqrt{(a+1)(1-d)}}\nonumber\\
&-&\frac{2 n}{\pi}\frac{\sqrt{-a d}}{\sqrt{(a+1)(1-d)}}\frac{a-1}{a+1}\bar{\Pi}'  \label{spn}
\end{eqnarray}
Equations (\ref{spikes}, \ref{spn}) can be used to find $a,d=a,d(n,\mathcal{S})$, which then using (\ref{ena}) gives the energy $\mathcal{E}=\mathcal{E}(n,\mathcal{S})$.
We checked numerically that the above equations give precisely the same energy for given $n, \mathcal{S}$ as the spiky string solution
from string theory \cite{k}.

Let us see how this matching works analytically. The spiky string solution \cite{k} is given by
\begin{equation}
\Delta \theta= \frac{\pi}{n}=\frac{\sinh 2\rho_0}{\sqrt{2} \sinh \rho_1}\frac{1}{\sqrt{u_1+u_0}}\bigg[\Pi[\frac{u_1-u_0}{u_1-1},p]-\Pi[\frac{u_1-u_0}{u_1+1},p]\bigg]  \label{spns}
\end{equation}
\begin{equation}
\mathcal{S}=\frac{n \cosh \rho_1}{\sqrt{2} \pi \sqrt{u_1+u_0}}\bigg[-(1+u_0) K[p] +(u_1+u_0) E[p] -\frac{u_0^2-1}{u_1+1}\Pi[\frac{u_1-u_0}{u_1+1},p]\bigg]  \label{bpns}
\end{equation}
\begin{equation}
\mathcal{E}-\omega \mathcal{S}=\frac{n \sqrt{u_1+u_0}}{\sqrt{2}\pi \sinh \rho_1}[K[p]-E[p]] \label{ansp}
\end{equation}
where $n$ is the number of the spikes and
\begin{equation}
u_0=\cosh 2 \rho_0, \quad \quad u_1= \cosh 2 \rho_1, \quad \quad \omega= \coth \rho_1, \quad \quad p=\frac{u_1-u_0}{u_1+u_0}
\end{equation}
$\rho_0$ and $\rho_1$ are the minimal and maximal values of the extension in $\rho$.

Using (\ref{deriv}) and (\ref{spikes}), equation (\ref{ena}) can be written as
\begin{equation}
\frac{\pi \Delta}{n}=\frac{\sqrt{-ad}}{\sqrt{(a+1)(1-d)}}\bigg[\frac{a+d}{a d}\frac{1}{\sqrt{-a d}-1}(\Pi- \sqrt{-a d} \ \bar{\Pi})+\frac{a+1}{a}K[r]+
\frac{(a+1)(1-d)}{2 a d }E[r]\bigg]
\end{equation}
Similarly, equation (\ref{spn}) can be rewritten as
\begin{eqnarray}
\frac{\pi \mathcal{S}}{n}&=&\frac{1}{\sqrt{(a+1)(1-d)}}\bigg[-\frac{1}{2}\frac{a+1}{a}(\sqrt{-a d}+a)K[r]+\frac{(a+1)(1-d)}{4 a d}(ad-\sqrt{-a d})E[r]\nonumber \\
&+&\frac{a+d + \sqrt{-a d} (\frac{1}{a}+\frac{1}{d})}{2 (\sqrt{-a d}-1)}[\sqrt{-ad} \ \bar{\Pi}-\Pi]  \label{ssl}
\end{eqnarray}
where
\beq
 \Pi= \Pi\left(\frac{\sinh2\rho_1-\sinh2\rho_0}{\cosh2\rho_1+\cosh2\rho_0},p\right), \ \ \
 \bar{\Pi}= \Pi\left(-\frac{\sinh2\rho_1-\sinh2\rho_0}{\cosh2\rho_1+\cosh2\rho_0},p\right)
\eeq
The main observation is that the combination
\begin{equation}
\frac{\pi}{n}\bigg(\mathcal{E}-\frac{\sqrt{-a d}+1}{\sqrt{-a d}-1} \mathcal{S}\bigg)=\frac{\sqrt{(a+1)(1-d)}}{\sqrt{-a d}-1}[K[r]-E[r]]
\end{equation}
contains no elliptic $\Pi$ functions.
Comparing to (\ref{ansp}) it is natural then to make the identification
\begin{equation}
e^{2 \rho_1}=\sqrt{- ad }, \quad \quad u_1=\frac{1-a d}{2 \sqrt{-a d }}
\end{equation}
Making further the identification
\begin{equation}
 e^{\rho_0} = \bigg(\frac{-d}{a}\bigg)^{1/4}, \quad \quad e^{\rho_1}=(-a d)^{1/4}
\end{equation}
we obtain
\begin{equation}
a= e^{2 \rho_1-2 \rho_0}, \quad \quad -d= e^{2 \rho_1+ 2 \rho_0}
\end{equation}
With these identifications we obtain the equations from the ABA
\begin{equation}
\frac{\pi}{n}[\mathcal{E}-\mathcal{S} \coth \rho_1]=\frac{1}{\sqrt{2}}\frac{\sqrt{u_0+u_1}}{\sinh \rho_1}[K[p]-E[p]], \quad \quad \frac{\pi}{n}=\frac{\sqrt{2}}{\sqrt{u_0+u_1} \sinh \rho_1}[\Pi-\bar{\Pi}]  \label{skp2}
\end{equation}
Also, we get that $r=p$.
In order to match equations (\ref{skp2}) along with the equation for $\mathcal{S}$ (\ref{ssl}) with the equations (\ref{spns}, \ref{bpns}, \ref{ansp}) we need the following identities
\begin{equation}
\Pi - \bar{\Pi}=\frac{\sinh 2 \rho_0}{2}\bigg(\Pi[\frac{u_1-u_0}{u_1-1},p]-\Pi[\frac{u_1-u_0}{u_1+1},p]\bigg)
\end{equation}
\begin{equation}
e^{\rho_1} \bar{\Pi} - e^{-\rho_1} \Pi= K[p] \sinh \rho_1 + \frac{\sinh 2 \rho_0}{2 \cosh \rho_1}\Pi[\frac{u_1-u_0}{u_1+1},p]
\end{equation}
We present a formal prof of these identities in Appendix A. We also checked them numerically. To conclude, we found that the spiky string solution for $\mathcal{J}=0$ obtained from ABA and classical string equations of motion perfectly match for all values of $\mathcal{S},n$.

\bigskip

An interesting open question is to find, from the ABA, the string solution in $AdS_3$ with spikes pointing inwards that was studied in \cite{ms}. The corresponding flat space solution has $1$ and $n+1$ left and right mode numbers, i.e. $n(x)$ for this solution is
\begin{eqnarray}
n(x)= \left\{
        \begin{array}{ll}
          -1, & d < x <c \\
          n+1, & b< x< a
        \end{array}
      \right.
\end{eqnarray}
As this solution \cite{ms} should be outside the $SL(2)$ sector it is not clear whether the $SL(2)$ ABA in the $\mathcal{J}\rightarrow 0$ limit can capture this solution. If this is not the case it would be interesting to find the BA that does capture this solution.

\renewcommand{\theequation}{5.\arabic{equation}}
 \setcounter{equation}{0}

\section{Spiky string with winding in $S^1 \subset S^5$}

In this section we find spiky string solutions with winding in $S^1\subset S^5$. Following \cite{kz} winding can be introduced by changing the momentum condition as
\begin{equation}
\int d x \frac{\tilde{\rho}(x)}{x}= - 2 \pi m
\end{equation}
As one can see from (\ref{reszero}, \ref{mom}) it is not possible to obtain a non-zero $m$ solution with $\tilde{\rho}(x)$ defined as in (\ref{rhosol}).
This is possible, however, for a different $\rho$ defined using a different $G$. More precisely we define a new $G$ (for convenience we denote this
different function again as $G$) as
\beq
 \Gg(w) = \frac{1}{\pi} \int_{C_1\cup C_2} \frac{F(w)}{F(x+i\epsilon)} \frac{\bar{n}(x)}{x-w} dx  \label{resolvent1}
\eeq
where
\begin{equation}
\bar{n}(x)= n(x)-\frac{2 \mathcal{J} x}{x^2-1}-\frac{2 m}{x^2-1} = \tilde{n}(x)-\frac{2 m}{x^2-1}
\end{equation}
Following the procedure in section 2 we introduce $\rho$ as
 \beq
 \Gg(x\pm i\epsilon) = \pm \bar{\rho}(x) + i \bar{n}(x), \ \ \ \ x\in C_1\cup C_1 , \ \ \ \epsilon\rightarrow 0^+.
\label{jumps2}
\eeq
with
\beq
\bar{\rho}(x) = \frac{1}{\pi} \ppint \left|\frac{F(x)}{F(x')}\right|\frac{\bar{n}(x')}{x'-x} dx'
\label{rhosol2}
\eeq
and we obtain the equation
\beqa
 \ppint_{C_1\cup C_2} \frac{\bar{\rho}(x')}{x-x'}(1-\frac{1}{x x'}) dx'=\pi (1-\frac{1}{x^2}) [\tilde{n}(x)-2 m - i G(0)]  \label{bas1}
\eeqa
where we have replaced for the residue at $w=0$. The condition for the residue at infinity ($w=\infty$) to vanish gives again the conditions
$G_0=G_1=0$. Of course all $G_i, G(0), H_1$ have to be computed now with $G$ as defined in (\ref{resolvent1}). We observe that
setting $-2m - i G(0)=0$ is indeed consistent with the momentum condition
\begin{equation}
\int d x \frac{\bar{\rho}(x)}{x}= i \pi G(0)= - 2 \pi m
\end{equation}
Therefore in this case $\bar{\rho}$ as defined in (\ref{rhosol2}) is a solution of equation (\ref{bas}), which we are indeed supposed to solve.

We can simplify (\ref{bas}) using $\int d x \frac{\bar{\rho}(x)}{x}= - 2 \pi m$ and obtain
\begin{equation}
\ppint_{C_1\cup C_2} \frac{\bar{\rho}(x')}{x-x'} dx'= \pi \bigg(n(x)-\frac{\mathcal{J}+m}{x-1}-\frac{\mathcal{J}-m}{x+1}\bigg)
\end{equation}
which is precisely the strong coupling equation found in \cite{kz}.

We consider two cuts with different lengths $[d,c]$ and $[b,a]$.  We take the cuts outside the $[-1,1]$ interval, i.e. $1<b<a$ and $d<c<-1$. Based on the corresponding solution in flat space, for spiky string with $n$ spikes we again take
\begin{eqnarray}
n(x)= \left\{
        \begin{array}{ll}
          -1, & d < x <c \\
          n-1, & b< x< a
        \end{array}
      \right.
\end{eqnarray}
The folded string with winding $m$ and angular momentum $\mathcal{J}$ is reached in the particular case with $n=2$. In contrast to the folded string solution, the cuts are no longer symmetrically distributed about the origin, and so the distribution of the Bethe roots is no longer symmetric, i.e. $\bar{\rho}(u) \neq \bar{\rho}(-u)$.

The resolvent $G(w)$ in the present case with non-trivial winding can be computed in a similar way as in section 4
\begin{eqnarray}
G(w)&=&- \frac{2 i n}{\pi \sqrt{(a-c)(b-d)}}\frac{F(w)}{(b-w)(w-c)}\bigg((b-c)\Pi[\frac{(a-b)(w-c)}{(a-c)(w-b)},r]+(w-b)K[r]\bigg)\nonumber\\
&-&i -2 i \mathcal{J}\bigg[\frac{w}{w^2-1}+\frac{F(w)}{2}\bigg(\frac{1}{F(1)(1-w)}-\frac{1}{F(-1)(1+w)}\bigg)\bigg]\nonumber\\
&-& 2 i m \bigg[\frac{1}{w^2-1}+\frac{F(w)}{2}\bigg(\frac{1}{F(1)(1-w)}+\frac{1}{F(-1)(1+w)}\bigg)\bigg]
\end{eqnarray}

Expanding $G(w)$ in large $w$ we obtain the coefficients
\begin{equation}
G_0=  \frac{2 i n}{\pi \sqrt{(a-c)(b-d)}}K[r] +i \mathcal{J}\bigg(\frac{1}{F(1)}+\frac{1}{F(-1)}\bigg)+ i m \bigg(\frac{1}{F(1)}-\frac{1}{F(-1)}\bigg)\label{als1}
\end{equation}
\begin{eqnarray}
G_1&=& -i  + \frac{2 i n}{\pi} \frac{b-c}{\sqrt{(a-c)(b-d)}}\Pi + \frac{2 i n}{\pi} \frac{c}{\sqrt{(a-c)(b-d)}} K[r] + i \mathcal{J}\bigg(\frac{1}{F(1)}-\frac{1}{F(-1)}\bigg)\nonumber\\
&+& i m \bigg(\frac{1}{F(1)}+\frac{1}{F(-1)}\bigg) \label{als2}
\end{eqnarray}
\begin{eqnarray}
G_2&=& -2 i \mathcal{J}+\frac{2 i n}{\pi}\frac{(b^2-c^2)}{\sqrt{(a-c)(b-d)}}\Pi + \frac{2 i n}{\pi}\frac{(b-c)^2 (a-b)}{(a-c)\sqrt{(a-c)(b-d)}}\Pi'\nonumber\\
&+& \frac{2 i n}{\pi}\frac{c^2}{\sqrt{(a-c)(b-d)}}K[r]
+i \mathcal{J}\bigg(\frac{1}{F(1)}+\frac{1}{F(-1)}\bigg)+ i m \bigg(\frac{1}{F(1)}-\frac{1}{F(-1)}\bigg)\nonumber\\
&-&\frac{i}{2}(a+b+c+d)   \label{als3}
\end{eqnarray}
Expanding $G(w)$ in small $w$, and also using the condition $G(0)= 2 i m$ lead to
\begin{equation}
\frac{i}{F(0)}= \frac{2 i n}{\pi}\frac{b-c}{b c \sqrt{(a-c)(b-d)}}\bar{\Pi} + \frac{i \mathcal{J}}{c}\bigg(\frac{1-c}{F(1)}+\frac{1+c}{F(-1)}\bigg)
+\frac{i m}{c} \bigg(\frac{1-c}{F(1)}-\frac{1+c}{F(-1)}\bigg)  \label{als4}
\end{equation}
\begin{eqnarray}
H_1&=& 2 i \mathcal{J}+\frac{i}{2}\bigg(\frac{1}{b}+\frac{1}{c}-\frac{1}{a}-\frac{1}{d}\bigg)- \frac{2 i n}{\pi}\frac{(b-c)^2 (a-b) c}{b^2 c^2 (a-c) b \sqrt{(a-c)(b-d)}}\bar{\Pi}' F(0)\nonumber\\
&-& \frac{i \mathcal{J}}{b c}F(0) \bigg(\frac{(1-c)(1-b)}{F(1)}+\frac{(1+c)(1+b)}{F(-1)}\bigg)\nonumber\\
&-& \frac{i m}{b c} F(0)\bigg(\frac{(1-c)(1-b)}{F(1)}-\frac{(1+c)(1+b)}{F(-1)}\bigg) \label{als5}
\end{eqnarray}
The spin and energy are again given by
\begin{equation}
\mathcal{E}-\mathcal{S}-\mathcal{J}=\frac{i}{2}H_1, \quad \quad \quad \mathcal{S}=-\frac{i}{4}(G_2+H_1)
\end{equation}
One needs to solve four equation, namely $G_0=G_1=0$ along with (\ref{als4}) and the spin equation for the unknown parameters $a,b,c,d$.
This is not possible analytically for arbitrary values of parameters. Below we consider the long string limit of the spiky string solution.

\subsection{Long string limit}

Let us consider the long string limit $a \rightarrow \infty$, $d \rightarrow -\infty$. More precisely let us consider the following scaling limit
\begin{equation}
d=- u a, \quad \quad a, \ \mathcal{J}, \ m \rightarrow \infty \ \ \ \ \texttt{with} \ \ \ \ \hat{\mathcal{J}}=\frac{\mathcal{J}}{A \ln a}, \ \ \hat{m}=\frac{m}{B \ln a}
\ \ \texttt{fixed}  \label{nscal}
\end{equation}
This is the generalization of the scaling found in \cite{ftt,bgk} to non-trivial winding. Expanding in large $a$ the equations (\ref{als1}, \ref{als2}, \ref{als4}) we obtain at leading order the following\footnote{We used the expansion of elliptic $\Pi$ integral for $\nu>\mu$

 $\Pi[\n,\m] \approx \sqrt{\frac{\n}{(1-\n)(\n-\m)}}\bigg(\frac{\pi}{2}- {\rm arcsin}
 \sqrt{\frac{1-\n}{1-\m}}\bigg)$.}
\begin{equation}
-\frac{d}{a}=u=\cot^2 \frac{\pi}{2 n}  \label{abc}
\end{equation}
\begin{equation}
(A \hat{\mathcal{J}}-B \hat{m}) \frac{1+c}{\sqrt{(-1-c)(1+b)}}+(A \hat{\mathcal{J}}+B \hat{m}) \frac{1-c}{\sqrt{(1-c)(b-1)}}-\frac{n}{\pi}=0 \label{lp1}
\end{equation}
\begin{equation}
(A \hat{\mathcal{J}}-B \hat{m}) \frac{1}{\sqrt{(-1-c)(1+b)}}+(A \hat{\mathcal{J}}+B \hat{m}) \frac{1}{\sqrt{(1-c)(b-1)}}-\frac{n}{\pi}=0 \label{lp2}
\end{equation}
Let us note that $u$ given in (\ref{abc}) does not depend on $m, \mathcal{J}$, and it is precisely the same as the corresponding relationship at $1$-loop weak coupling BA (see eq. 3.17 in \cite{fkt}).
Equations (\ref{lp1}, \ref{lp2}) can be solved for $b,c$
\begin{equation}
b=\frac{4 \pi^2 A B \hat{\mathcal{J}}\hat{m}+\sqrt{n^4+16 \pi^4 A^2 B^2 \hat{\mathcal{J}^2}\hat{m}^2+4 \pi^2 n^2 (A^2 \hat{\mathcal{J}}^2+B^2 \hat{m}^2)}}{n^2}
\end{equation}
\begin{equation}
c=\frac{4 \pi^2 A B \hat{\mathcal{J}}\hat{m}-\sqrt{n^4+16 \pi^4 A^2 B^2 \hat{\mathcal{J}^2}\hat{m}^2+4 \pi^2 n^2 (A^2 \hat{\mathcal{J}}^2+B^2 \hat{m}^2)}}{n^2}
\end{equation}
 Expanding (\ref{als3}, \ref{als5}) in large $a$ we obtain
\begin{equation}
 \mathcal{S}=\frac{a n}{4 \pi \sqrt{u}}+...
\end{equation}
\begin{equation}
\mathcal{E}-\mathcal{S}= \frac{\ln a}{2 \sqrt{-b c}}\bigg(\frac{n}{\pi}+ \frac{4 \pi}{n}(A^2 \hat{\mathcal{J}}^2 + B^2 \hat{m}^2)\bigg) + ...
\end{equation}
which plugging the expressions for $b,c$ give the following result
\begin{equation}
\mathcal{E}-\mathcal{S}=\frac{n}{2 \pi} \sqrt{1+ \frac{4 \pi^2}{n^2}\bigg(\frac{\mathcal{J}^2}{\ln^2 \mathcal{S}}+\frac{m^2}{\ln^2 \mathcal{S}}\bigg)}\ln \mathcal{S}+... \label{scal1}
\end{equation}
The particular case with $n=2$ corresponds to the folded spinning string with winding.

\subsection{Matching the string solution}

A general ansatz for classical rigid string solutions in $AdS_5 \times S^1$ that includes both angular momentum and winding in $S^1$ was considered in \cite{iktt}. The solution can be written in terms of a set of equations, which are to be solved. Let us recall the equations obtained in \cite{iktt}.
Following the notation in \cite{iktt} let us consider the following rigid string
ansatz \cite{iktt}
\bea
&&Y_0=r_0(u)e^{i \varphi_0(u)+i w_0 \tau}\ , \qquad \quad Y_1=r_1(u)e^{i
\varphi_1(u)+i w_1 \tau}\ ,
\qquad \quad X=e^{i \psi(u)+ i \nu \tau}\ ,\\
&& \quad \quad u\equiv
\sigma+\beta \tau, \ \ \ \ \ \ \ \ r_0^2- r^2_1=1
\eea
where $Y_r$ are the embedding coordinates.
The conformal constraints give the equation of motion for $r_1$
\begin{equation}
(\beta^2-1)^2
r_1^{'2}=(1+r_1^2)\bigg(\frac{C_0^2}{1+r_1^2}+ w_0^2
(1+r_1^2)-\frac{C_1^2}{r_1^2}- w_1^2 r_1^2 -D^2-
\nu^2\bigg) \label{r1}
\end{equation}
and the condition
\begin{equation}
\omega_0 C_0 + \omega_1 C_1 + D \nu=0
\end{equation}
where $C_0, C_1,D$ are constants obtained by integrating the equations of motion for $\varphi_0, \varphi_1, \psi$ (see \cite{iktt} for details).
Introducing the variable $v$ as
\begin{equation}
v= \frac{1}{1+ 2 r_1^2}=\frac{1}{\cosh 2 \rho}, \qquad \qquad 0 \leq v \leq 1
\end{equation}
the equation of motion for $v$, (\ref{r1}),  becomes
\begin{equation}\la{vew}
v'= \frac{\sqrt{2 v P(v)}}{1-\beta^2}
\end{equation}
where
\begin{equation}
P(v)=a (v-v_1)(v-v_2)(v-v_3), \ \ \ \ \ \ \ \ \ a \equiv -4 C_0^2 - 4 C_1^2 +2 (D^2+ \nu^2) - w_0^2- w_1^2=\frac{w_1^2 -w_0^2}{v_1 v_2 v_3}  \label{qho}
\end{equation}
$v_n$ are the three roots of $P(v)=0$. It turns out that for the solution of interest $P(v)$ has two positive roots $0 \leq v_2 \leq v_3 \leq 1 $, and one negative $v_1 \leq 0$, thus the physical motion takes place in the range $v_2 \leq v \leq v_3$. The expressions for the constants $C_0,C_1$ in term of $v_1,v_2,v_3$ are
\begin{equation}
C_0^2= \frac{w_0^2-w_1^2}{8}\frac{(1+v_1)(1+v_2) (1+v_3)}{v_1 v_2 v_3}, \quad
C_1^2= \frac{w_0^2-w_1^2}{8}\frac{(1-v_1)(1-v_2)(1-v_3)}{v_1 v_2 v_3}  \label{cis}
\end{equation}
To get  solutions with  $n$ spikes we need to
glue together a number of $2 n$ pieces of integrals between a
minimum ($v_2$) and a maximum  ($v_3$). In other words, wherever it appears,
the integral $\int d u$ is to be replaced by
\begin{equation}
\int d u = 2n \int_{v_2}^{v_3} \frac{d v }{v'}= \frac{2 n (1-\beta^2)}{\sqrt{-2 a}}I_1
\end{equation}
where we define the integrals $I_n$ below.

The winding number $m$ in $S^1 \subset S^5$, can be written as\footnote{For convenience here we take $m\rightarrow -m$ as compared to the same formula in \cite{iktt}.}
\begin{equation}
m = \frac{D-\beta \nu }{\pi \sqrt{-2 a}} n I_1  \label{sir}
\end{equation}
while the condition of no winding in the $t$ direction gives the condition
\begin{equation}
2 C_0 I_5 +  w_0 \beta I_1=0  \label{cak}
\end{equation}
For solutions with $n$ spikes the conserved charges are
\begin{equation}
\frac{\mathcal{ \pi E}}{ n}=\frac{\beta C_0}{\sqrt{-2 a}}I_1 + \frac{w_0}{2 \sqrt{-2 a}}I_3, \quad \frac{\pi \mathcal{S} }{
n}=-\frac{\beta C_1}{\sqrt{-2 a}}I_1 + \frac{w_1}{2 \sqrt{-2 a}}I_2, \quad \frac{\pi \mathcal{J} }{ n}=\frac{\nu - \beta D}{\sqrt{-2 a}}I_1  \label{aiu}
\end{equation}
Also, the number of spikes can be introduced as the angle between a minimum and a maximum, and can be expressed as
\begin{equation}
\frac{\pi}{n} = -\frac{2}{\sqrt{-2 a}}(C_1 I_6 + \frac{w_1}{w_0}C_0 I_5)  \label{nsp}
\end{equation}
The relevant integrals that appear in the above formulas are
\begin{equation}
I_1= \int_{v_2}^{v_3} d v \frac{1}{\sqrt{-v (v-v_1)(v-v_2)(v-v_3)}}=\frac{2}{\sqrt{v_3 (v_2-v_1)}}K[s],
\end{equation}
\begin{equation}
I_2= \int_{v_2}^{v_3} d v\frac{1-v}{v\sqrt{-v (v-v_1)(v-v_2)(v-v_3)}}= -\frac{2}{v_1 v_2}\sqrt{\frac{v_2-v_1}{v_3}} E [s]
+\frac{2 (\frac{1}{v_1}-1)}{\sqrt{v_3 (v_2-v_1)}}K[s]
\end{equation}
\begin{equation}
I_3= \int_{v_2}^{v_3} d v\frac{1+v}{v\sqrt{-v (v-v_1)(v-v_2)(v-v_3)}}=-\frac{2}{v_1 v_2}\sqrt{\frac{v_2-v_1}{v_3}} E [s]+ \frac{2 (\frac{1}{v_1}+1)}{\sqrt{v_3 (v_2-v_1)}}K[s]
\end{equation}
\begin{equation}
I_5= \int_{v_2}^{v_3} d v \frac{v}{(1+v) \sqrt{-v (v-v_1)(v-v_2)(v-v_3)}}=\frac{2 v_2}{(1+v_2)\sqrt{v_3(v_2-v_1)}}\Pi [\frac{v_3-v_2}{v_3(1+v_2)},s]
\end{equation}
\begin{equation}
I_6=\int_{v_2}^{v_3} d v \frac{v}{(1-v) \sqrt{-v (v-v_1)(v-v_2)(v-v_3)}}=\frac{2 v_2}{(1-v_2)\sqrt{v_3(v_2-v_1)}}\Pi [\frac{v_3-v_2}{v_3(1-v_2)},s]
\end{equation}
where
\begin{equation}
s= \frac{v_1 (v_2-v_3)}{v_3 (v_2-v_1)}
\end{equation}

It would be interesting to precisely match the set of equations describing these solutions from the string side for arbitrary charges with the corresponding equations from the BA obtained in section 5. We have checked this matching explicitly for $m=\mathcal{J}=0$ in section 4. For non-zero $\mathcal{J},m$ we study below the string solution in the scaling limit when the string touches the boundary and $\mathcal{S}, \mathcal{J}, m$ are all large.

\bigskip

More precisely the scaling (\ref{nscal}) is obtained in the limit $v_2 \rightarrow 0$ with $D, \nu$ finite. We observe that in this limit $I_{1,2,3}$ diverge but $I_{4,5}$ stay finite. To have a well defined polynomial $P(v)$ in this limit we need $\omega_0 \rightarrow \omega_1$. This implies $C_{0,1}$ are finite and so is $\beta I_1$. As in the BA case, let us introduce the finite, fixed quantities
\begin{equation}
\hat{\mathcal{J}}=\frac{\mathcal{J}}{A \ln \frac{1}{v_2}}, \ \ \ \ \ \ \ \ \ \ \hat{m}=\frac{m}{B \ln \frac{1}{v_2}}
\end{equation}
In the scaling limit the energy, spins and winding can be written as
\begin{equation}
\frac{\pi (\mathcal{E}-\mathcal{S})}{n}=\frac{\omega_1}{\sqrt{-2 a} \sqrt{-v_1 v_3}}\ln \frac{1}{v_2}+..., \ \ \ \ \ \ \
\frac{\pi \mathcal{S}}{n}=\frac{\omega_1}{\sqrt{-2 a} \sqrt{-v_1 v_3}} \ \frac{1}{v_2}+...
\end{equation}
\begin{equation}
\frac{\pi}{n} A \hat{\mathcal{J}}=\frac{\nu}{\sqrt{-2 a} \sqrt{-v_1 v_3}}+..., \ \ \ \ \ \ \ \  \ \frac{\pi}{n} B \hat{m}=\frac{D}{\sqrt{-2 a} \sqrt{-v_1 v_3}}+...
\end{equation}
In the limit the definition of spikes (\ref{nsp}) gives a relationship between $v_1, v_3$
\begin{equation}
\cos \frac{\pi}{n}= \frac{v_3 \sqrt{1-v_1^2}+v_1 \sqrt{1-v_3^2}}{v_1-v_3}
\end{equation}
Using the equations (\ref{qho}, \ref{cis}) in the limit we obtain $2 \sqrt{-2 a} \sqrt{-v_1 v_3}=\sqrt{\omega_1^2-D^2-\nu^2}$. Plugging in the equations for charges we obtain the energy in the scaling limit
\begin{equation}
\mathcal{E}-\mathcal{S}=\frac{n}{2 \pi} \sqrt{1+ \frac{4 \pi^2}{n^2}\bigg(\frac{\mathcal{J}^2}{\ln^2 \mathcal{S}}+\frac{m^2}{\ln^2 \mathcal{S}}\bigg)}\ln \mathcal{S}+...
\end{equation}
which precisely matches the equation obtained from the ABA (\ref{scal1}) in the same scaling limit.

\section*{Acknowledgments }

We are grateful  to A. Belitsky, L. Freyhult, R. Roiban and A.  Tseytlin  for useful discussions and correspondence. This work was supported in part by NSF under grant PHY-0805948, DOE under grant DE-FG02-91ER40681 and by the Alfred P. Sloan foundation.

\def \bi {\bibitem}

\renewcommand{\theequation}{A.\arabic{equation}}
 \setcounter{equation}{0}

\appendix
\subsection*{Appendix A:  On the relationships among elliptic integrals}

In the main text we made use of the identities
\begin{equation}
\Pi - \bar{\Pi}=\frac{\sinh 2 \rho_0}{2}\bigg(\Pi[\frac{u_1-u_0}{u_1-1},p]-\Pi[\frac{u_1-u_0}{u_1+1},p]\bigg)
\label{A1eq}
\end{equation}
\begin{equation}
e^{\rho_1} \bar{\Pi} - e^{-\rho_1} \Pi= K[p] \sinh \rho_1 + \frac{\sinh 2 \rho_0}{2 \cosh \rho_1}\Pi[\frac{u_1-u_0}{u_1+1},p]
\label{A2eq}
\end{equation}
where
\beq
 \Pi= \Pi\left(\frac{\sinh2\rho_1-\sinh2\rho_0}{\cosh2\rho_1+\cosh2\rho_0},p\right), \ \ \
 \bar{\Pi}= \Pi\left(-\frac{\sinh2\rho_1-\sinh2\rho_0}{\cosh2\rho_1+\cosh2\rho_0},p\right)
\eeq
and
\beq
p=\frac{u_1-u_0}{u_1+u_0}
\eeq
 and we remind the reader of the notation $u_0=\cosh2\rho_0$, $u_1=\cosh2\rho_1$. To prove these identities, we define the function
\beq
 F(\rho_0,\rho_1) = \Pi - \half K[p] - \frac{e^{\rho_1} \sinh(2\rho_0)}{4\sinh\rho_1} \Pi[\frac{u_1-u_0}{u_1-1},p] +
               \frac{e^{\rho_1} \sinh(2\rho_0)}{4\cosh\rho_1} \Pi[\frac{u_1-u_0}{u_1+1},p]
\eeq
It is easily seen that, if $F(\rho_0,\rho_1)=0$ and $F(-\rho_0,-\rho_1)=0$ then the desired results (\ref{A1eq}), (\ref{A2eq}) follow. So we need to prove
that $F$ vanishes identically. Using the properties of the elliptic integrals one can compute the derivatives
\beqa
 \frac{\partial F}{\partial u_1} &=& \frac{e^{2\rho_1}+ u_0}{u_1+u_0} \frac{F}{2\sinh 2\rho_1} \\
 \frac{\partial F}{\partial u_0} &=& \frac{F}{2(u_1+u_0)}
\eeqa
 These equations can be integrated to give
\beq
 F = A e^{\rho_1} \sqrt{u_0+u_1}
\eeq
 where $A$ is a constant independent of $\rho_{0},\rho_{1}$. It is straight-forward to verify that when $\rho_0=\rho_1$ we have $F(\rho_0,\rho_0)=0$
which implies that $A=0$ and therefore proves the identity $F=0$.

 Having proved that $F(\rho_0,\rho_1)=0$ it is interesting to take the limit $\rho_0\rightarrow 0$ and derive the result
\beq
\Pi(q,q^2)=\half K(q^2) + \frac{\pi}{4(1-q)}
\eeq
As a check this last result is also easily proved by defining
\beq
G(q) =  \Pi(q,q^2)-\half K(q^2)
\eeq
and showing that
\beq
\frac{\partial G(q)}{\partial q} = \frac{G(q)}{1-q}
\eeq
which implies
\beq
 G(q) = \frac{B}{1-q}
\eeq
Using $\Pi(0,0)=K(0)=\frac{\pi}{2}$ we get $G(0)=\frac{\pi}{4}$ which implies $B=\frac{\pi}{4}$.



\end{document}